\documentclass[12pt]{article}
\pdfoutput=1

\usepackage{amsmath}
\usepackage{amsfonts}
\usepackage{amssymb}

\usepackage[
      colorlinks=true,
      linkcolor=blue,
      urlcolor=blue,
      filecolor=black,
      citecolor=red,
      pdfstartview=FitV,
      pdftitle={},
        pdfauthor={ Michael Gutperle, John D. Miller},
        pdfsubject={},
        pdfkeywords={},
        pdfpagemode={},
        bookmarksopen=true
      ]{hyperref}

\usepackage{color}
\usepackage{graphicx}

\usepackage{sectsty}
\sectionfont{\large}

\renewcommand{\thefootnote}{\fnsymbol{footnote}}
\renewcommand{\thanks}[1]{\footnote{#1}}
\newcommand{\starttext}{
\setcounter{footnote}{0}
\renewcommand{\thefootnote}{\arabic{footnote}}}
\newcommand{\bea}{\begin{eqnarray}}
\newcommand{\eea}{\end{eqnarray}}
\newcommand{\ee}{\end{equation}}
\newcommand{\be}{\begin{equation}}

\numberwithin{equation}{section}

\DeclareMathOperator{\tr}{tr}
\usepackage{ dsfont } 

\long\def\symbolfootnote[#1]#2{\begingroup%
\def\thefootnote{\fnsymbol{footnote}}\footnote[#1]{#2}\endgroup}

\begin{document}
\setlength{\baselineskip}{18pt}

\starttext
\setcounter{footnote}{0}

\begin{flushright}
\today
\end{flushright}

\bigskip

\begin{center}

{\Large \bf   A note on entanglement entropy for topological interfaces in RCFTs}

\vskip 0.4in

{\large   Michael Gutperle and John D. Miller}

\vskip .2in

{ \it Department of Physics and Astronomy }\\
{\it University of California, Los Angeles, CA 90095, USA}\\[0.5cm]

\bigskip
\href{mailto:gutperle@physics.ucla.edu}{\texttt{gutperle}}\texttt{, }
\href{mailto:johnmiller@physics.ucla.edu}{\texttt{johnmiller@physics.ucla.edu}}

\bigskip

\bigskip

\end{center}
 
\begin{abstract}

\setlength{\baselineskip}{18pt}

In this paper we calculate the entanglement entropy for topological interfaces in rational conformal field theories for the case where the interface lies at the boundary of the entangling interval and for the case where it is located in the center of the entangling interval. We compare the results to each  other and also to the recently calculated left/right entropy of a related BCFT.
We also comment of the entanglement entropies for topological interfaces  for a free compactified boson and Liouville theory.

\end{abstract}

\setcounter{equation}{0}
\setcounter{footnote}{0}


\newpage

\section{Introduction}
\label{sec1}

In two dimensional conformal field theories  the construction  of conformal interfaces between $CFT_1$ and $CFT_2$ is equivalent to  the  construction of conformal boundary conditions in the folded CFT which is the tensor product $CFT_1\otimes \overline{CFT}_2$ \cite{Oshikawa:1996dj,Bachas:2001vj}. There is a special class of interfaces which are  called topological interfaces. In the folded CFT the condition for an interface to be topological is \cite{Bachas:2001vj}
\begin{align}
\Big(L^{(1)}_n- \bar L^{(2)}_{-n}\Big)\mid    B\rangle=0, \quad \quad \Big(L^{(2)}_n- \bar L^{(1)}_{-n}\Big) \mid  B\rangle=0
\end{align}
This condition implies that the interface is completely transmissive and these interfaces are  called topological since their position and shape can be deformed without cost of energy.  In addition one can define a fusion product of topological interfaces by bringing two of them close together \cite{Bachas:2007td}. It has been argued  in \cite{Frohlich:2004ef,Frohlich:2006ch,Bachas:2008jd} that topological interfaces  can furnish spectrum generating symmetries. Topological interfaces have been constructed for a single free boson in \cite{Bachas:2001vj,Fuchs:2007tx} and for $n$ free bosons compactified on an $n$ dimensional  torus in 
\cite{Bachas:2012bj}. Topological interfaces in orbifold theories have been studied in \cite{Brunner:2013ota}.

The entanglement entropy of a spatial region ${\cal A}$ is defined by the von Neumann entropy of a reduced density matrix which is obtained by integrating out the degrees of freedom localized in the complement of ${\cal A}$. For two dimensional CFTs  \cite{Holzhey:1994we,Calabrese:2004eu} the entanglement entropy, where ${\cal A}$ is a single interval of length $L$, is given by
\begin{align}\label{sadef}
{\cal S}_{\cal A}= {c\over 3} \ln \left( { L\over \epsilon} \right) + \ln g
\end{align}
Here $\epsilon$ is an UV cutoff and we dropped terms which vanish as $\epsilon\to 0$. The logarithmically divergent term only depends on the central charge of the CFT.  Generically the constant term is non-universal and dependent on the definition of the cutoff since a simple rescaling of the cutoff is equivalent to adding a constant  term to the entanglement entropy.

In the presence of a boundary or an interface the constant term becomes physical as one can study the difference in entanglement entropy for a system with an interface and without, and the dependence on the UV cutoff cancels in this case. In the case of a conformal interface there are two different geometric setups which one can consider. First if the interface is located at the center of the entangling region ${\cal A}$ then \cite{Calabrese:2004eu} the entanglement entropy is of the form given in \eqref{sadef} and the constant term is the so called $g$-factor or boundary entropy \cite{Oshikawa:1996dj}. A second geometry is given by choosing the interface to be located at the boundary of the entangling region ${\cal A}$. The entanglement entropy in this case has only been calculated for some specific cases, notably for generic conformal interfaces for a  single free boson in   \cite{Sakai:2008tt}
 and the Ising model in \cite{Brehm:2015lja}. The entanglement entropy in this case takes the form
 \begin{align}\label{satohres}
 {\cal S}_{\cal A}= {c\over 6}  f({\cal I}) \ln \left( {L\over \epsilon} \right) + \ln g 
 \end{align}
where $ f({\cal I})$ is a complicated function depending on the parameters of the conformal interface, which we schematically denote by ${\cal I}$. The entanglement entropies for Janus like interface theories \cite{Bak:2003jk,Bak:2007jm,Chiodaroli:2009yw} have been calculated for the symmetric geometry in \cite{Azeyanagi:2007qj,Chiodaroli:2010ur} and for the case where the interface is at the boundary of the entangling surface in \cite{Gutperle:2015hcv}.

The goal of this note is to calculate the entanglement entropies for both geometric setups for topological interfaces in rational CFTs. The structure of the paper is as follows: In section \ref{sec2} we review the construction of topological interfaces in rational CFTs which goes back to the work of Petkova and Zuber \cite{Petkova:2000ip}. In section \ref{sec3} we adapt the calculation of Sakai and Satoh  \cite{Sakai:2008tt} to calculate the entanglement entropy for the case where the topological interface is at the boundary of the entangling surface.  In section \ref{sec4} we adapt the calculation of Sakai and Satoh to re-derive the entanglement entropy \eqref{sadef} for the symmetric interface and also give an argument that the location of the interface does not change the result as long as it is a finite distance away from the boundaries of the entangling interval.  In section \ref{sec5} we compare the entanglement entropies for various specific cases. We also include the recently computed left/right entropy  \cite{PandoZayas:2014wsa,Das:2015oha} for reference.  In section  \ref{sec6} we compare results  of entanglement entropies for compact bosons which were previously obtained in the literature.  In section \ref{sec7} we use the construction of topological interfaces in Liouville theory given in    \cite{Sarkissian:2009aa,Poghosyan:2015oua} to attempt a calculation of  the entanglement entropy for this system. We close with a discussion of future directions of research in section \ref{sec8}.

\medskip

{\bf Note added}: While this paper was finalized a paper \cite{brunner} appeared, which has significant overlap with the material presented  in section \ref{sec3}.

\section{Topological interfaces in RCFT}
\label{sec2}
In this section we  consider the construction   of topological interfaces in rational CFTs.
A rational CFT (RCFT) \cite{Moore:1988uz} contains a finite number of primary states and hence a finite number  of representations of the Virasoro algebra, labeled by $i$, with characters $\chi_i(q)$.  The partition function on the torus is given by
\begin{align}
Z= \sum_{i\bar j} Z_{i\bar j} \; \chi_i(q)\chi_{\bar j}(\bar q)
\end{align}
where $Z_{i\bar j} $ are positive integers which denote how many times a representation appears in the spectrum of the theory. We mostly limit ourselves the the case where  $Z_{i\bar j}=\delta_{i\bar j}$ and the theory has a diagonal spectrum.  

The canonical examples for rational CFTs\footnote{We limit ourselves to minimal models with respect to the Virasoro algebra here, generalizing the discussion to rational CFTs with respect to extended conformal algebras would be very interesting.} are the unitary minimal models which have central charge
\begin{align}
c=1-{6\over m(m+1)}, \quad \quad m=3,4,\cdots
\end{align}
and the primaries are labeled by two integers $r=1,2,\cdots,m-1$ and $s=1,2,\cdots r$ and have conformal dimension
\begin{align}
h_{r,s}={\big((m+1)r-m s\big)^2 -1\over 4 m(m+1)}
\end{align}
The simplest minimal model is the Ising model which has $m=3$ and hence  we have $c={1\over 2}$ and three primaries with $h=0$, $h={1\over 2}$  and $h={1\over 16}$.

In \cite{Petkova:2000ip,Petkova:2001ag} twisted partition functions for rational CFTs were studied. They are characterized by the insertion of an operator $I$  into the partition function, where $I$  satisfies 
\begin{align}\label{xdef}
[L_n,I]=[\bar L_n,I]=0
\end{align}
In \cite{Petkova:2000ip}   a classification of such  operators was given analogous to the construction of Cardy states \cite{Cardy:1989ir}. For the diagonal theories one finds 
\begin{align}\label{interfxa}
I_a= \sum_{i} {S_{ai}\over S_{0i}} P^{i\bar i}
\end{align}
Here $P^{i\bar i}$ is a projector on the space spanned by the $i-th$ primary and its descendants
\begin{align}
P^{i\bar i}= \sum_{n\bar n} |i,n\rangle \otimes |i, \bar n\rangle \langle i,n| \otimes \langle i, \bar n|
\end{align}
This means that in the simple diagonal case there are as many topological interfaces as there are primaries, where for simplicity we assume that each primary only appears once in the theory; a degeneracy can be easily included in the construction.\footnote{See \cite{Fuchs:2002cm} for a discussion of more general projectors including non-diagonal theories.}
The matrix  $S$ is the modular $S$ matrix which denotes how the characters transform under modular transformation $q=e^{2\pi i \tau}$
\begin{align}
\chi(-{1\over \tau}) = \sum_j S_{ij} \chi_j(\tau)
\end{align}
The conjugate interface operator is given by
\begin{align}
I^\dagger_a= \sum_{i} \left({S_{ai}\over S_{0i}}\right)^* P^{i\bar i}
\end{align}
In summary it is notable that the classification of  \cite{Petkova:2000ip,Petkova:2001ag}  of twisted partition function also provides us with a classification of topological interfaces in RCFTs.

\section{Entanglement entropy at a topological interface}
\label{sec3}

The entanglement entropy for an interval  ${\cal A}$ is given by  the von Neumann entropy of the reduced density  matrix $\rho_{\cal A}={\rm tr}_{\bar{\cal A}} |0\rangle\langle 0|$, where $\bar {\cal A}$ is the complement of ${\cal A}$.  The replica trick relates the entanglement entropy to the  Renyi entropies as follows
 \begin{align}\label{renyien}
 {\cal S}_{\cal A} &= -{\partial \over \partial K} {\rm tr}  \left.\rho_{\cal A}^K \right|_{K=1}
 \end{align}

 \begin{figure}[!t]
  \centering
  \includegraphics[width=115mm]{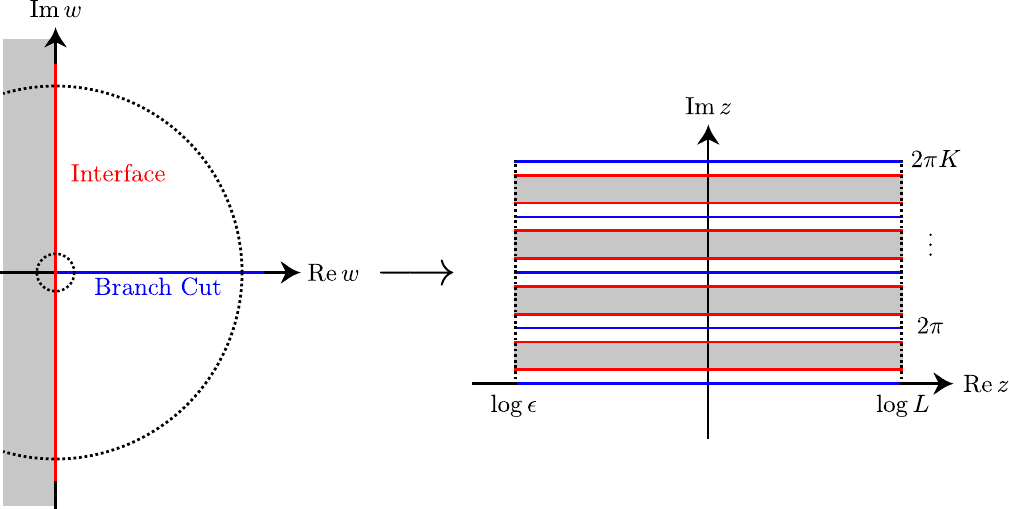}
  \caption{The  logarithmic map $z=\ln w$ maps the K-sheeted Riemann surface to  the geometry on the right. The circles on the left part of the figure  correspond to an  UV cutoff located  $|w|=\epsilon$ and an  IR cutoff located at $|w|=L$. This figure was adapted from \cite{Brehm:2015lja}.}
  \label{fig:replica}
\end{figure}

 The Renyi entropies are calculated  by a path integral over a K-sheeted Riemann surface with  branch cuts running along ${\cal A}$. The entanglement entropy can then be derived from the partition function  $Z(K)$ on the K-sheeted Riemann surface by 
  \begin{align}\label{entanglea}
 {\cal S}_{\cal A}&= (1-\partial_K) \left.\ln Z(K)\right|_{K=1}
 \end{align} 
In \cite{Sakai:2008tt}  it has been argued that the interface can be included in  the K-sheeted partition function $Z(K)$. The interface gets mapped via $z=\ln w$ to a covering space  (see figure \ref{fig:replica}).
Introducing an UV  cutoff $\epsilon$ and an IR cutoff $L$ and imposing periodic boundary conditions for simplicity, the K-th replica partition function can be expressed as a trace 
 \begin{align}\label{zkformula}
 Z(K) = \tr \Big( I  e^{- t H} I^\dagger e^{-tH} \Big)^K
 \end{align} 
 where  the ``time" is related to the cutoffs by
 \begin{align}\label{defta}
 t={2\pi^2 \over  \ln(L/\epsilon)}
 \end{align}  and 
 \begin{align}
 H&= L_0+\bar L_0-{c\over 12}  
 \end{align}
 is the Hamiltonian of the CFT. The $K$-th partition function with a topological  interface \eqref{interfxa} labeled by a primary $a$ inserted is 
\begin{align}\label{zafirst}
Z_a(K)&=\tr\Big( \Big[ I_a q^{L_0-{c\over 24}}\bar q^{\bar L_0-{c\over 24}} I_a^\dagger  q^{L_0-{c\over 24}}\bar q^{\bar L_0-{c\over 24}}\Big]^K\Big)\nonumber\\
&= \tr\Big( [ I_a I_a^\dagger ]^K  q^{2K (L_0-{c\over 24})} \bar q^{2K(\bar L_0-{c\over 24})}\Big)\nonumber\\ 
&= \sum_i \left| {S_{ai}\over S_{0i} }\right|^{2K}  \chi_i(q^{2K})\chi_{\bar i}(\bar q^{2K})
\end{align}
where we have introduced $q=\bar q=e^{-t}$. In the second line we have used \eqref{xdef} to commute $I_a$ through the Hamiltonian and in the third line we used the fact that the $P^{ii}$ in \eqref{interfxa} are projectors to the $i$-th representation and the trace produces the associated character $\chi_i$.

Since we are interested in taking the UV cutoff $\epsilon\to 0$ (and  taking  $L\to \infty$), we have to evaluate \eqref{zafirst} in the limit $q\to 1$. With the identification of a new modular parameter $\tau'$ by
\begin{align}\label{tauiden}
q^{2K}=e^{-2K t}=e^{2\pi i \tau'}
\end{align}
with $t$ given in \eqref{defta}, 
the limit can  taken  by performing a modular transformation on the characters
\begin{align}\label{tlimit}
\lim_{q\to 1}  \chi_i(q^{2K})\chi_{\bar i}(\bar q^{2K})
&=\lim_{\tau'\to 0}  \chi_i(\tau')\chi_i(\bar{\tau}') \nonumber \\
&= \lim_{\tau'\to 0}  \sum_{j,k} S_{ij}S^*_{ik}\, \chi_j(-1/\tau')\,\chi_k(-1/\bar{\tau}')\nonumber\\
&=\sum_{j,k} S_{ij}S^*_{ik}e^{\pi^2 c\over  6 Kt}e^{-{2\pi^2  h_j\over Kt}}e^{-{2\pi^2 h_k\over Kt}}\left(1+o[e^{-{2\pi^2/ K t} }]\right)
\end{align}
In the limit $t\rightarrow 0$ the leading  contribution  in \eqref{tlimit} will come from the vacuum characters which have  $h_j=h_k=0$. In that case the partition function \eqref{zafirst} becomes
\begin{equation}
Z_a(K) \approx \exp\left(\frac{c}{12 K}\ln\frac{L}{\epsilon}\right)\sum_i|S_{ai}|^{2K}|S_{0i}|^{2-2K}+ \cdots
\end{equation}
where the dots indicate terms which vanish as the cutoff is taken to zero. Further calculating
\begin{align}
\big(1-\partial_K\big)\ln\left(\sum_i|S_{ai}|^{2K}|S_{0i}|^{2-2K}\right)|_{K=1} &= -2\frac{\sum_i|S_{ai}|^{2K}|S_{0i}|^{2-2K}\left(\ln|S_{ai}|-\ln|S_{0i}|\right)}{\sum_j|S_{aj}|^{2K}|S_{0j}|^{2-2K}}|_{K=1}\nonumber\\
&= -2 \sum_i|S_{ai}|^2\ln\left|\frac{S_{ai}}{S_{0i}}\right|
\end{align}
where we have repeatedly used the fact  that $S$ is symmetric, unitary, and  in particular the relation ${\sum_j|S_{aj}|^2}=1$. Putting everything together we arrive at the following expression for the entanglement entropy at a topological interface
\begin{equation}\label{safinal}
{\cal S}_a=\frac{c}{6}\ln\frac{L}{\epsilon}-2\sum_i|S_{ai}|^2\ln\left|\frac{S_{ai}}{S_{0i}}\right|
\end{equation}

\section{Symmetric and left/right entanglement entropy}
\label{sec4}

For an interface which is located symmetrically on the entangling interval ${\cal A}$ the entanglement entropy has been calculated by \cite{Calabrese:2004eu} and is given by

\begin{align}\label{symmee}
{\cal S}^{{\rm symm}} = {c\over 3} \ln {L\over \epsilon}+ \ln g_B
\end{align}
where $g_B$ is the boundary g-factor which when the interface is folded into a boundary state is  determined by the overlap of the  boundary state corresponding to the doubled interface with the vacuum state \cite{Oshikawa:1996dj, Harvey:1999gq}
\begin{align}
g_B= \langle 0\mid B\rangle
\end{align}
For the topological interface \eqref{interfxa} the boundary state becomes 
\begin{align}
\mid B_a\rangle = \sum_{i} {S_{ai}\over S_{0i}} \sum_{n\bar n} |i,n\rangle \otimes |i, \bar n\rangle \mid  i,\bar n\rangle \otimes \mid i,   n\rangle 
\end{align}
Consequently the $g$ factor is given by
\begin{align}
g_B= {S_{a0}\over S_{00}} 
\end{align}
and the symmetric entanglement entropy becomes
\begin{align}\label{ssymfinal}
{\cal S}_a^{{\rm symm}} = {c\over 3} \ln {L\over \epsilon}+ \ln \left({S_{a0}\over S_{00}} \right)
\end{align}

\begin{figure}[!t]
  \centering
  \includegraphics[width=130mm]{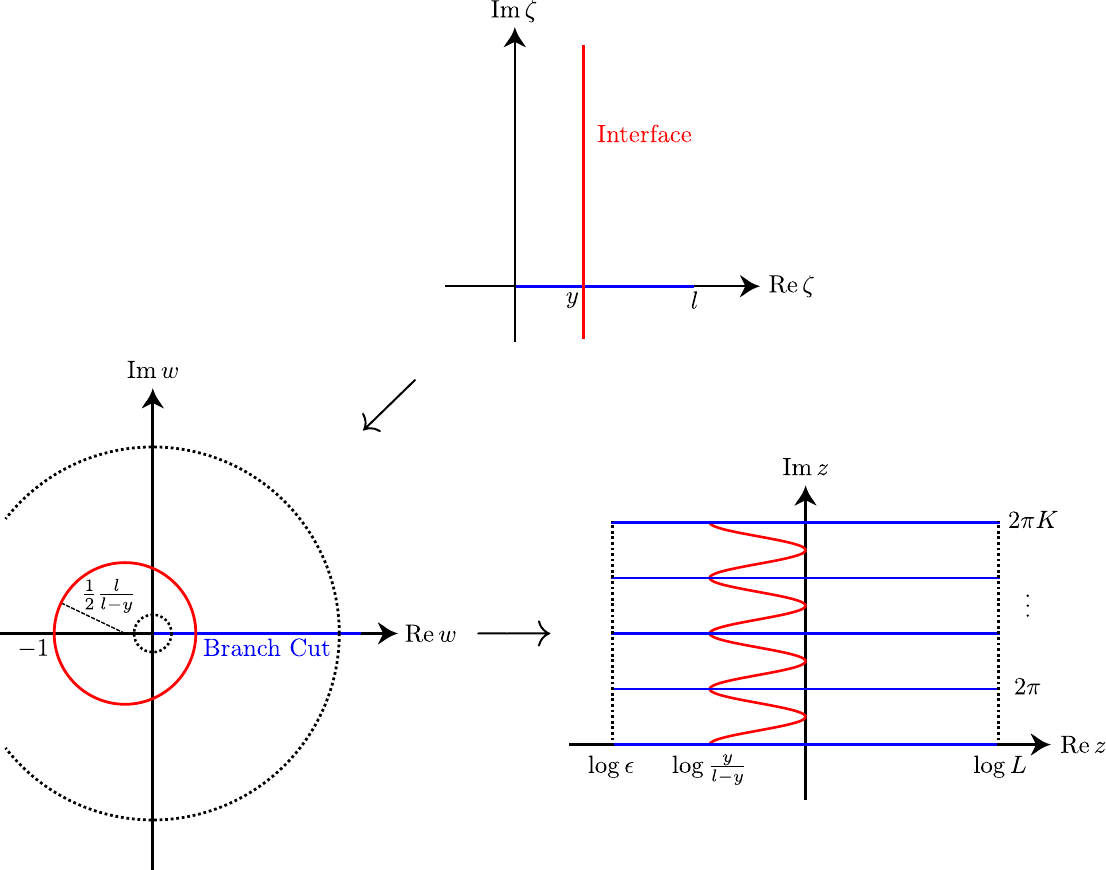}
  \caption{Mapping of a non-central interface to the replica torus. }
  \label{fig:figure2}
\end{figure}

Up to now we have considered the symmetric case where the interface is located at the center of the entangling interval ${\cal A}$. There is however a simple argument showing that for topological interfaces the location of the interface does not change the result as long as it is a finite distance away from the boundary of the entangling interval. We illustrate the argument in figure \ref{fig:figure2}. We start in the $\zeta$ plane with a finite  interval ${\cal A}$ with boundary at $\zeta=0$ and $\zeta=l$, where the interface is located along $\zeta= y+ i \xi, \xi \in R$. We map the $\zeta$ plane into the $w$ plane by the map
\begin{align}
z= {\zeta\over l-\zeta } 
\end{align}

This maps the finite interval to the positive real axis and the interface $I$ gets mapped to an off centre  circle. Finally we perform the replica map to the $z$ coordinate via $z=\ln w$  and impose periodic boundary conditions as before at the cutoff $z=\ln \epsilon$ and $z=\ln L$. This produces again a torus. Unlike the case of the interface at the boundary here the interface is mapped into a vertical curve on the torus. For a topological interface it is clear that the shape can be changed and changing the location along the real part of $z$ corresponds to changing the original location $y$ of the interface. This shows that partition function on the $K$-th sheeted Riemann surface is independent of $y$ as long as the interface is a finite distance away form the cutoffs.
We can be more specific and evaluate the partition function
\begin{align}\label{zsymmb}
Z_a(K)&= \tr(  I_a e^{-t H} )\nonumber\\
&= \sum_i {S_{ai}\over S_{0i}} \chi_i(q) \chi_i (\bar q)
\end{align}
where $q=\bar q= e^{-{\pi^2\over K t}}$ where $t$ is again given by \eqref{defta}, hence in the limit of vanishing cutoff the sum over representations in the partition function gets projected on the vacuum character and one has
\begin{align}
Z_a(K)\sim {S_{a0}\over S_{00}} \exp\left(\frac{c}{12 K}\ln\frac{L}{\epsilon}\right) +\cdots
\end{align}
Applying the replica formula \eqref{entanglea} one obtains
\begin{align}\label{symresx}
{\cal S}_a^{\rm symm} ={c\over 6} \ln\left(\frac{L}{\epsilon} \right) + \ln \left({S_{a0}\over S_{00}}\right)
\end{align}
Comparing (\ref{symresx}) with  (\ref{ssymfinal}) one notices an extra factor of $1/2$ in (\ref{symresx}) in the $\ln(L/\epsilon)$ term. This seeming discrepancy comes from the fact that the replica calculation leading to (\ref{symresx}) calculates the entanglement entropy for a semi-infinite entangling surface (as we take $L$ to be very large) with only one end  point, whereas the result of Cardy and Calabrese  (\ref{ssymfinal}) is for an interval with two end  points, which doubles the logarithmically divergent contribution according to the area law for entanglement entropy. The same remark applies when one compares (\ref{sadef})  and (\ref{safinal}).

Additionally it is clear that for a topological interface moving the interface along the real axis in the $z$ coordinates does not change \eqref{zsymmb} as the interface operator commutes with the generator of these translations, which is the Hamiltonian. 
It is clear from Figure \ref{fig:figure2} that the independence of the symmetric entanglement entropy from the location of the interface breaks down if the interface approaches  the UV cutoff $\epsilon$, as part of the interface would  be removed by the cutoff. This explains why the entanglement entropies (\ref{safinal}) and (\ref{ssymfinal}) can be different.

A third type of entanglement entropy which takes a similar form is the so called left/right entanglement entropy \cite{PandoZayas:2014wsa,Das:2015oha,Schnitzer:2015gpa}. This is defined for a boundary CFT, where the entanglement entropy is calculated with a reduced density matrix obtained by tracing over left-moving modes. Interestingly for a boundary CFT defined by a  Cardy state \cite{Cardy:1989ir} (for a single copy of the CFT, not the doubled one we are considering in the previous sections)
\begin{align}
\mid B^{\rm Cardy}_a\rangle&= \sum_j {S^j_a\over \sqrt{S_0^j}} \mid j\rangle\rangle
\end{align}
where $\mid j\rangle\rangle$ are the Ishibashi states \cite{Ishibashi:1988kg} enforcing conformal
boundary conditions. We quote the result of the calculation of the left/right entanglement entropy which is also labeled by a primary $a$ in a RCFT, obtained in \cite{Das:2015oha}

 \begin{align}\label{slrfinal}
{\cal S}^{\rm l/r}_ {a} = {\pi c l\over 24 \epsilon}- \sum_j S_{aj}^2 \ln \left( {S_{aj}^2\over S_{0j}}\right)
\end{align}
The physical interpretation of the left/right  entanglement entropy (as it is non-geometrical) is not clear at this point as well as its relation to the other two entropies is not clear at the moment. The similarities of the resulting entropies might still suggest that such a relation exists. A better understanding of the relation of the cutoffs utilized may be necessary to accomplish this. 
 
\section{Examples of entanglement entropies}
\label{sec5}

For the $m$-th unitary minimal models the modular $S$ matrix is given by  (see e.g. \cite{DiFrancesco:1997nk}
).
\begin{align}
S_{rs; \rho\sigma}= 2\sqrt{2\over m(m+1)} (-1)^{1+ s \rho+ r \sigma} \sin\left( \pi {m+1\over m} r \rho\right)\sin\left( \pi {m\over m+1} s\sigma\right)
\end{align}
Using this formula it is in principle straightforward  to evaluate the three entanglement entropies ${\cal S}_a$ given in \eqref{safinal}, ${\cal S}_a^{{\rm symm}}$ given in \eqref{ssymfinal}
 and ${\cal S}^{\rm l/r}_ {a} $ given in \eqref{slrfinal}. Here we   give a table for the two simplest cases, namely the Ising model with $m=3$ and the tri-critical Ising model with $m=4$.
 
 The Ising model has 3 primaries which we  can label by the conformal dimension $h=0,{1\over 16} , {1\over 2}$ and the $S$ matrix becomes
 \begin{align}
 S&={1\over 2}
\left(
\begin{array}{ccc}
 1 & 1  & \sqrt{2}   \\
 1 &  1 & -\sqrt{2}  \\
\sqrt{2}  &-\sqrt{2}   &0   
\end{array}
\right)
 \end{align}
 The entanglement entropies then take the following values
 \begin{table}[htp]
\begin{center}
\caption{entanglement entropies for the Ising model}
\begin{tabular}{|c|c|c|c|}
\hline
& ${\cal S}_a$ &${\cal S}^{{\rm symm}}_a$ & ${\cal S}^{\rm l/r}_a$ \\
\hline
\hline
$h=0$ &0&0& ${3\over 4} \ln2$\\
\hline
$h={1\over 2}$&0&0&$ {3 \over 4} \ln 2$ \\
\hline
$h={1\over 16}$&$ -\ln 2$ &${\ln 2\over 2}$  &0\\
\hline
\end{tabular}
\end{center}
\label{Table1}
\end{table}

 The next simplest minimal model is the tri-critical Ising model which has $m=4$ and has six primary states which are labelled by their conformal dimension 
 \begin{align}
 h=0,{1\over 10} ,{3\over 5},{3\over 2} ,{3\over 80}, {7\over 16}
 \end{align}
 The modular S-matrix is given by
 \begin{align}
 S&=
\left(
\begin{array}{cccccc}
 s_2 & s_1  & s_1&s_2 & \sqrt{2}s_1 & \sqrt{2} s_2  \\
s_1 & -s_2  & -s_2&s_1 & \sqrt{2}s_2 & -\sqrt{2} s_1  \\
s_1 & -s_2  & -s_2&s_1 & -\sqrt{2}s_2 & \sqrt{2} s_1  \\
s_2 & s_1  & s_1&s_2 & -\sqrt{2}s_1 & -\sqrt{2} s_2  \\
\sqrt{2}s_1&\sqrt{2} s_2&-\sqrt{2} s_2&-\sqrt{2} s_1&0&0\\
\sqrt{2}s_2&-\sqrt{2} s_1&\sqrt{2} s_1&-\sqrt{2} s_2&0&0\\
\end{array}
\right)
 \end{align}
 where $s_1$ and $s_2$ are given by
 \begin{align}
 s_1= \sin \left({2\pi\over 5}\right), \quad \quad s_2= \sin \left({4\pi\over 5}\right)
 \end{align}
The entanglement entropies then take the following values
\begin{tiny}
  \begin{table}[htp]
\begin{center}
\caption{entanglement entropies for the tri-critical Ising model}
\begin{tabular}{|c|c|c|c|}
\hline
a& ${\cal S}_a$ &${\cal S}^{{\rm symm}}_a$ & ${\cal S}^{\rm l/r}_a$ \\
\hline
\hline
$h=0$ &0&0& $ {-2 \sqrt{5} \coth^{-1} \sqrt{5}+ \ln{32768\over 3125}\over 4}$\\
\hline
$h= {1\over 10}$&$-\sqrt{5} \coth^{-1} \left({3\over \sqrt{5}}\right)$&${1\over 2} \ln {3+\sqrt{5}\over 2}$ &${15 \ln2 -5 \ln 5+ \sqrt{5}\ln (9-4 \sqrt{5}) \over 4}$ \\
\hline
$h= {3\over 5}$&$-\sqrt{5} \coth^{-1} \left({3\over \sqrt{5}}\right)$&${1\over 2} \ln {3+\sqrt{5}\over 2}$ &${15 \ln2 -5 \ln 5+ \sqrt{5}\ln (9-4 \sqrt{5}) \over 4}$ \\
\hline
$h={3\over 2}  $&0&0& $ {-2 \sqrt{5} \coth^{-1} \sqrt{5}+ \ln{32768\over 3125}\over 4}$\\
\hline
$h= {3\over 80}$&${(-5+\sqrt{5}) \ln(3-\sqrt{5})-(5+\sqrt{5}) \ln(3+\sqrt{5})\over 2}$ &${1\over 2} \ln (3+\sqrt{5})$& ${-5\ln 5+\sqrt{5} \ln(9-4 \sqrt{5})\over 4}$ \\
\hline
$h= {7\over 16}$&$-5 \ln 2$&$ {\ln 2\over 2}$&${-2 \sqrt{5} \coth^{-1} \sqrt{5} - 5 \ln 5\over 4}$\\
\hline
\end{tabular}
\end{center}
\label{Table2}
\end{table}
\end{tiny}

\section{Entanglement entropies for a compact boson}
\label{sec6}

A general class of conformal interfaces for a free boson was constructed in \cite{Bachas:2001vj}.
The interfaces are characterized by two radii $R_1,R_2$ and two relatively prime integers $k_1,k_2$. 
Note that a compact boson is not a RCFT and the results of the previous sections cannot be directly applied. Instead we will review results obtained in the literature to contrast the entanglement entropies for  this case.  

These correspond to an interface between two compactified bosons where the compactification radius jumps from $R_1$ to $R_2$ across the interface. In the doubled BCFT description the interface corresponds to a geometric D1 brane stretched on a rectangular torus with radii $R_1$ and $R_2$ where its one dimensional world volume wraps $k_1$ times around the  $R_1$ circle and $k_2$ times around the $R_2$ circle.
In general these interfaces are not topological but for special values of the radii where the following condition is satisfied
\begin{align}\label{topcon}
k_1 R_1=k_2 R_2
\end{align}
the defects become topological. It was shown in  \cite{Sakai:2008tt}  that the complicated behavior of the logarithmically  divergent term of the entanglement entropy for the interface at the boundary of the entangling surface \eqref{satohres} simplifies and the entanglement entropy becomes
\begin{align}
{\cal S}= {1\over 6}   \ln \left( {L\over \epsilon} \right) + \ln |k_1 k_2|
\end{align}
We can contrast this explicit expression    to the  one of symmetric entanglement entropy, the $g_B$ factor for a general interface is given by 
\cite{Bachas:2001vj,Bachas:2007td} 
\begin{align}
g_B = \sqrt{k_1^2 R_1^2 + k_2^2 R_2^2\over 2 R_1 R_2}
\end{align}
and hence the symmetric entanglement entropy for the topological interface which satisfies \eqref{topcon} becomes
\begin{align}
{\cal S}^{\rm symm} = {1\over 3}   \ln \left( {L\over \epsilon} \right) + {1\over 2} \ln |k_1 k_2|
\end{align}
Where the constant part is half the value of the constant part of ${\cal S}$.  Note that if we consider the case of an interface between identical CFTs, we have $R_1=R_2$ and by \eqref{topcon} $k_1=k_2$ for topological interfaces.
\section{ Remarks on entanglement entropies for Liouville theory}
\label{sec7}

 In \cite{Sarkissian:2009aa,Drukker:2010jp,Poghosyan:2015oua} topological  interfaces for the Liouville CFT (see \cite{Teschner:2001rv,Nakayama:2004vk} for  reviews with references to the original  literature) were constructed following the procedure which was used  for RCFTs.  There are two types of defects 
 which are both of the form
  \begin{align}\label{louint}
 I= \int_{Q/2+i P} d\alpha \; D(P ){\cal P}^{\alpha}
 \end{align}
 where we integrate $P$ over the positive real line, i.e. $P\in (0,\infty)$, and one has  $Q=b+1/b$, which determines the central charge as $C=1+ 6Q^2$.
 Here ${\cal P}$ is a projector on the continuum of primary states  labeled by $P$ and their descendants.
 \begin{align}
 {\cal P}^{\alpha}= \sum_{M,N}|\alpha ,M\rangle \otimes |\overline{ \alpha, N}\rangle \langle \alpha,M| \otimes \langle  \overline{\alpha , N}|
 \end{align}
 As shown in  \cite{Sarkissian:2009aa} one can distinguish the two $D$ by associating them with the discrete degenerate primary states labeled by two positive integers
 \begin{align}\label{disccase}
 D_{m,n}(P)=  {\sinh ({2\pi m P\over b}) \sinh( 2\pi n b P) \over  \sinh ({2\pi  P\over b}) \sinh( 2\pi  b P)}
 \end{align}
 and a non-degenerate primary state labeled by a continuous real parameter $s$
 \begin{align}\label{contcase}
 D_s(P)=  {\cos (4 \pi P s )\over 2 \sinh (2\pi b P) \sinh({2\pi P\over b})} 
 \end{align}
 We can now calculate the $K$-sheeted partition function \eqref{zkformula}
with the interface  \eqref{louint} inserted.
Using the fact that the projectors satisfy
\begin{align}
{\cal P}^\alpha {\cal P}^\beta = \delta(\alpha-\beta){\cal P}^\alpha
\end{align}
and the fact that the interface operator $I$ satisfies  \eqref{xdef} we arrive at
\begin{align}
Z(K)= \int _{Q/2+i P} dP \; \big(D(P)\big)^{2K} \chi_P(q^{2K} )\chi_P({\bar q}^{2K})
\end{align}
where $\chi_P(q)$ is the character of the non-degenerate Liouville primary field labeled by $P$ and is given by
\begin{align}\label{chardefa}
\chi_P(\tau) ={ q^{P^2} \over q^{1\over 24} \prod_{n=1}^\infty(1-q^n)}
\end{align}
where $q=e^{2\pi i\tau}$
and we can use the following formula for the modular transformation of the character \eqref{chardefa}
\begin{align}
\chi_P(-{1\over \tau}) =\sqrt{2} \int_{-\infty}^\infty dP' \chi_{P'} (\tau) e^{4 \pi i P P'} 
\end{align}
With the identification \eqref{tauiden}
and  $t$ given by \eqref{defta} as before, the modular transformed $K$-sheeted partition function becomes
\begin{align}
Z(K)= 2 \int_0^\infty  dP\; \big(D(P)\big)^{2K} \int_{-\infty}^\infty   dP' e^{4 \pi i P P'}  \chi_{P'} ( i {\pi \over Kt} )   \int_{-\infty}^\infty  d\bar P' e^{4 \pi i P \bar P'}  \chi_{\bar P'} ( i {\pi \over Kt} )  
\end{align}
In the limit $t\to 0$ we can replace the full character $\chi_P(q)$ by its leading term $q^{P^2-{1\over 24}}$ and perform the gaussian integrals over $P'$ and ${\bar P}'$ which produce the same result. Hence we arrive at
\begin{align}\label{zkliou}
Z(K)= { K t \over 4\pi} e^{\pi^2 \over 6 K t} \int_0^\infty  dP  \big(D(P)\big)^{2K}  e^{-4 P^2 K t} + \cdots
\end{align}
where the dots denote terms which vanish as   $t$ goes to zero. We would now like to use this expression to calculate the entanglement entropy using the replica formula \eqref{entanglea}.
Note that for the case where $D$ is labeled by a continuous parameter $s$ and given by \eqref{contcase}  $\big(D(P)\big)^{2K}$ in the integral \eqref{zkliou} vanishes for large $P$. It is therefore legitimate to drop the exponent $e^{-4 P^2 K t}$  in the integral and the non vanishing terms in entanglement entropy for this case  are given by
\begin{align}\label{Ssresult}
{\cal S}_s={1\over 6} \ln {L\over \epsilon} +(1-\partial_K)\Big( \ln  { K t \over 4\pi} \int_0^\infty  dP  \big(D_s(P)\big)^{2K} \Big)\mid_{K=1}
\end{align}
We notice two curious features of this result. First, the logarithmically divergent term is multiplied by ${1\over 6}$ which is what one would expect for  a $c=1$ CFT, whereas the central charge of the Liouville theory is given by $C_L=1+6Q^2$. 
 A possible explanation for this behavior lies in the fact that for the interface labelled by (\ref{contcase}) only the continuous primaries with conformal dimension $\Delta=Q^2/4+P^2$ appear. Hence the vacuum with $\Delta=0$ is excluded and the factor of $1/3$ in front of (\ref{Ssresult}) is most likely associated with a shifted effective central charge. 

 Second, apart from finite terms as $t\to 0$ we also obtain an additonal divergent term of the form $\ln(\ln {L\over \epsilon} )$ from the second term \eqref{Ssresult}. The significance and interpretation of this term is not clear at this point and a more careful treatment of the cutoff might be necessary.
For the interfaces labeled by discrete integers $m,n$ defined in \eqref{disccase}, $D_{mn}$ diverges for large $P$ and the full integral has to be evaluated first in order to obtain the entanglement entropy. We leave this problem for future work.

\section{Discussion}
\label{sec8}

In this paper we have discussed entanglement entropies  in the presence of topological defects in two geometric settings, namely when the interface is located at the boundary of the entangling interval ${\cal A}$ and when it is in the center of the entangling interval.
For topological defects in  RCFTs the logarithmic part of the entanglement entropy is always universal (this is not the case for general conformal interfaces) and the constant term can be expressed in a compact form in terms of the modular matrix $S$. Note that the  entanglement entropies have a similar form in terms of the modular matrix S as the recently obtained left/right entanglement entropy for a related BCFT, but the physical relation of the  left/right entanglement entropy to the others is not clear at the moment.

There are several directions in which our results can be generalized. We have limited ourselves to RCFTs with diagonal partition functions. The construction of \cite{Petkova:2000ip} also includes non-diagonal theories and it would be interesting to understand the entanglement entropy for this case. We also only considered CFTs which are rational with respect to the Virasoso algebra, it would also be very interesting to repeat the analysis for RCFTs with respect to extended chiral algebras.

Since the large $m$ limit of minimal models is conjectured to approach a non-rational $c=1$ CFT which is different from a free boson \cite{Runkel:2001ng} it would be interesting to study the continuation  of the minimal model entanglement entropy.

It would also be interesting to understand the entanglement entropy for Liouville theory better. Apart from the calculations sketched in section \ref{sec7} one might also consider semiclassical limits where $b\to 0$ and analyze the role of topological defects in classical Liouville theory following  \cite{Aguirre:2013zfa,Poghosyan:2015oua}.

 \section*{Acknowledgements}
  This work was supported in part by National Science Foundation grant PHY-13-13986. The work of MG  was in part supported by a fellowship of the Simons Foundation. MG thanks the Institute for Theoretical Physics, ETH Z\" urich, for hospitality while part of this work was performed.


\newpage


  \begin{thebibliography}{99}
 
\bibitem{Oshikawa:1996dj}
  M.~Oshikawa and I.~Affleck,
  ``Boundary conformal field theory approach to the critical two-dimensional Ising model with a defect line,''
  Nucl.\ Phys.\ B {\bf 495} (1997) 533
  [cond-mat/9612187].



\bibitem{Bachas:2001vj}
  C.~Bachas, J.~de Boer, R.~Dijkgraaf and H.~Ooguri,
  ``Permeable conformal walls and holography,''
  JHEP {\bf 0206} (2002) 027
  [hep-th/0111210].


\bibitem{Bachas:2007td}
  C.~Bachas and I.~Brunner,
  ``Fusion of conformal interfaces,''
  JHEP {\bf 0802} (2008) 085
  [arXiv:0712.0076 [hep-th]].
 
 
 
 
 
 
\bibitem{Frohlich:2004ef}
  J.~Frohlich, J.~Fuchs, I.~Runkel and C.~Schweigert,
  ``Kramers-Wannier duality from conformal defects,''
  Phys.\ Rev.\ Lett.\  {\bf 93} (2004) 070601
  doi:10.1103/PhysRevLett.93.070601
  [cond-mat/0404051].
 
\bibitem{Frohlich:2006ch}
  J.~Frohlich, J.~Fuchs, I.~Runkel and C.~Schweigert,
  ``Duality and defects in rational conformal field theory,''
  Nucl.\ Phys.\ B {\bf 763} (2007) 354
  doi:10.1016/j.nuclphysb.2006.11.017
  [hep-th/0607247].
 
 
\bibitem{Bachas:2008jd}
  C.~P.~Bachas,
  ``On the Symmetries of Classical String Theory,''
  arXiv:0808.2777 [hep-th].
 
 
 
\bibitem{Fuchs:2007tx}
  J.~Fuchs, M.~R.~Gaberdiel, I.~Runkel and C.~Schweigert,
  ``Topological defects for the free boson CFT,''
  J.\ Phys.\ A {\bf 40} (2007) 11403
  [arXiv:0705.3129 [hep-th]].
 
\bibitem{Bachas:2012bj}
  C.~Bachas, I.~Brunner and D.~Roggenkamp,
  ``A worldsheet extension of O(d,d:Z),''
  JHEP {\bf 1210} (2012) 039
  doi:10.1007/JHEP10(2012)039
  [arXiv:1205.4647 [hep-th]].
 

 
 
\bibitem{Brunner:2013ota}
  I.~Brunner, N.~Carqueville and D.~Plencner,
  ``Orbifolds and topological defects,''
  Commun.\ Math.\ Phys.\  {\bf 332} (2014) 669
  [arXiv:1307.3141 [hep-th]].
 

 
 
\bibitem{Holzhey:1994we}
  C.~Holzhey, F.~Larsen and F.~Wilczek,
  ``Geometric and renormalized entropy in conformal field theory,''
  Nucl.\ Phys.\ B {\bf 424} (1994) 443
  doi:10.1016/0550-3213(94)90402-2
  [hep-th/9403108].
 
 
\bibitem{Calabrese:2004eu}
  P.~Calabrese and J.~L.~Cardy,
  ``Entanglement entropy and quantum field theory,''
  J.\ Stat.\ Mech.\  {\bf 0406} (2004) P06002
  doi:10.1088/1742-5468/2004/06/P06002
  [hep-th/0405152].
 
  
\bibitem{Sakai:2008tt}
  K.~Sakai and Y.~Satoh,
  ``Entanglement through conformal interfaces,''
  JHEP {\bf 0812} (2008) 001
  [arXiv:0809.4548 [hep-th]].
  


\bibitem{Brehm:2015lja}
  E.~M.~Brehm and I.~Brunner,
  ``Entanglement entropy through conformal interfaces in the 2D Ising model,''
  JHEP {\bf 1509} (2015) 080
  [arXiv:1505.02647 [hep-th]].

\bibitem{Petkova:2000ip}
  V.~B.~Petkova and J.~B.~Zuber,
  ``Generalized twisted partition functions,''
  Phys.\ Lett.\ B {\bf 504} (2001) 157
  [hep-th/0011021].
 

\bibitem{Bak:2003jk}
  D.~Bak, M.~Gutperle and S.~Hirano,
  ``A Dilatonic deformation of AdS(5) and its field theory dual,''
  JHEP {\bf 0305} (2003) 072
  [hep-th/0304129].

 
\bibitem{Bak:2007jm}
  D.~Bak, M.~Gutperle and S.~Hirano,
  ``Three dimensional Janus and time-dependent black holes,''
  JHEP {\bf 0702} (2007) 068
  [hep-th/0701108].


\bibitem{Chiodaroli:2009yw}
  M.~Chiodaroli, M.~Gutperle and D.~Krym,
  ``Half-BPS Solutions locally asymptotic to AdS(3) x S**3 and interface conformal field theories,''
  JHEP {\bf 1002} (2010) 066
  [arXiv:0910.0466 [hep-th]].


 
  
\bibitem{Azeyanagi:2007qj}
  T.~Azeyanagi, A.~Karch, T.~Takayanagi and E.~G.~Thompson,
  ``Holographic calculation of boundary entropy,''
  JHEP {\bf 0803} (2008) 054
  [arXiv:0712.1850 [hep-th]].

  
\bibitem{Chiodaroli:2010ur}
  M.~Chiodaroli, M.~Gutperle and L.~Y.~Hung,
  ``Boundary entropy of supersymmetric Janus solutions,''
  JHEP {\bf 1009} (2010) 082
  [arXiv:1005.4433 [hep-th]].

  
\bibitem{Gutperle:2015hcv}
  M.~Gutperle and J.~D.~Miller,
  ``Entanglement entropy at holographic interfaces,''
  arXiv:1511.08955 [hep-th].
  

   

 
 
\bibitem{PandoZayas:2014wsa}
  L.~A.~Pando Zayas and N.~Quiroz,
  ``Left-Right Entanglement Entropy of Boundary States,''
  JHEP {\bf 1501} (2015) 110
  [arXiv:1407.7057 [hep-th]].
 
\bibitem{Das:2015oha} 
  D.~Das and S.~Datta,
  ``Universal features of left-right entanglement entropy,''
  Phys.\ Rev.\ Lett.\  {\bf 115}, no. 13, 131602 (2015)
  [arXiv:1504.02475 [hep-th]].
 
 
\bibitem{Sarkissian:2009aa}
  G.~Sarkissian,
  ``Defects and Permutation branes in the Liouville field theory,''
  Nucl.\ Phys.\ B {\bf 821} (2009) 607
  [arXiv:0903.4422 [hep-th]].
  
\bibitem{Drukker:2010jp}
  N.~Drukker, D.~Gaiotto and J.~Gomis,
  ``The Virtue of Defects in 4D Gauge Theories and 2D CFTs,''
  JHEP {\bf 1106} (2011) 025
  doi:10.1007/JHEP06(2011)025
  [arXiv:1003.1112 [hep-th]].
  
\bibitem{Poghosyan:2015oua}
  H.~Poghosyan and G.~Sarkissian,
  ``On classical and semiclassical properties of the Liouville theory with defects,''
  JHEP {\bf 1511} (2015) 005
  [arXiv:1505.00366 [hep-th]].

 
 \bibitem{brunner}
 E.M. Brehm, I. Brunner, D. Jaud and  C. Schmidt-Colinet,
``Entanglement and topological interfaces", arxiv:1512.05945[hep-th]
 
 
\bibitem{Moore:1988uz}
  G.~W.~Moore and N.~Seiberg,
  ``Polynomial Equations for Rational Conformal Field Theories,''
  Phys.\ Lett.\ B {\bf 212} (1988) 451.
 
 
 
\bibitem{Petkova:2001ag}
  V.~B.~Petkova and J.~B.~Zuber,
  ``The Many faces of Ocneanu cells,''
  Nucl.\ Phys.\ B {\bf 603} (2001) 449
  doi:10.1016/S0550-3213(01)00096-7
  [hep-th/0101151].
 
 
 
\bibitem{Cardy:1989ir}
  J.~L.~Cardy,
  ``Boundary Conditions, Fusion Rules and the Verlinde Formula,''
  Nucl.\ Phys.\ B {\bf 324} (1989) 581.
  doi:10.1016/0550-3213(89)90521-X
 
\bibitem{Fuchs:2002cm}
  J.~Fuchs, I.~Runkel and C.~Schweigert,
  ``TFT construction of RCFT correlators 1. Partition functions,''
  Nucl.\ Phys.\ B {\bf 646} (2002) 353
  doi:10.1016/S0550-3213(02)00744-7
  [hep-th/0204148].
 
\bibitem{Harvey:1999gq}
  J.~A.~Harvey, S.~Kachru, G.~W.~Moore and E.~Silverstein,
  ``Tension is dimension,''
  JHEP {\bf 0003} (2000) 001
  doi:10.1088/1126-6708/2000/03/001
  [hep-th/9909072].
  
  
\bibitem{Schnitzer:2015gpa}
  H.~J.~Schnitzer,
  ``Left-Right Entanglement Entropy, D-Branes, and Level-rank duality,''
  arXiv:1505.07070 [hep-th].
  
\bibitem{Ishibashi:1988kg}
  N.~Ishibashi,
  ``The Boundary and Crosscap States in Conformal Field Theories,''
  Mod.\ Phys.\ Lett.\ A {\bf 4} (1989) 251.
  
\bibitem{DiFrancesco:1997nk}
  P.~Di Francesco, P.~Mathieu and D.~Senechal,
  ``Conformal Field Theory,'', Springer 
  
  
  
\bibitem{Teschner:2001rv}
  J.~Teschner,
  ``Liouville theory revisited,''
  Class.\ Quant.\ Grav.\  {\bf 18} (2001) R153
  [hep-th/0104158].
  
  
\bibitem{Nakayama:2004vk}
  Y.~Nakayama,
  ``Liouville field theory: A Decade after the revolution,''
  Int.\ J.\ Mod.\ Phys.\ A {\bf 19} (2004) 2771
  [hep-th/0402009].
  
  
\bibitem{Runkel:2001ng}
  I.~Runkel and G.~M.~T.~Watts,
  ``A Nonrational CFT with c = 1 as a limit of minimal models,''
  JHEP {\bf 0109} (2001) 006
  [hep-th/0107118].
 
 
\bibitem{Aguirre:2013zfa}
  A.~R.~Aguirre,
  ``Type-II defects in the super-Liouville theory,''
  J.\ Phys.\ Conf.\ Ser.\  {\bf 474} (2013) 012001
    [arXiv:1312.3463 [math-ph]].
 
  \end{thebibliography}
\end{document}